\documentclass
[aps,prb,superscriptaddress,reprint,showpacs,amssymb,amsmath,floatfix,citeautoscript,longbibliography]{revtex4-1}

\usepackage{graphicx} 
\usepackage{dcolumn} 
\usepackage{bm} 

\usepackage{ulem}
\usepackage[]{color}

\begin{document}

\title{Impact of Fe substitution on the electronic structure of URu$_2$Si$_2$ }

\author{Martin~Sundermann}
 \affiliation{Institute of Physics II, University of Cologne, Z{\"u}lpicher Stra{\ss}e 77, 50937 Cologne, Germany}
 \affiliation{Max Planck Institute for Chemical Physics of Solids, N{\"o}thnitzer Stra{\ss}e 40, 01187 Dresden, Germany}
\author{Andrea~Amorese}
 \affiliation{Institute of Physics II, University of Cologne, Z{\"u}lpicher Stra{\ss}e 77, 50937 Cologne, Germany}
 \affiliation{Max Planck Institute for Chemical Physics of Solids, N{\"o}thnitzer Stra{\ss}e 40, 01187 Dresden, Germany}
\author{Daisuke~Takegami}
 \affiliation{Max Planck Institute for Chemical Physics of Solids, N{\"o}thnitzer Stra{\ss}e 40, 01187 Dresden, Germany}
\author{Hlynur~Gretarsson}
 \affiliation{Max Planck Institute for Chemical Physics of Solids, N{\"o}thnitzer Stra{\ss}e 40, 01187 Dresden, Germany}
 \affiliation{PETRA III, Deutsches Elektronen-Synchrotron (DESY), Notkestra{\ss}e 85, 22607 Hamburg, Germany}
\author{Hasan~Yava\c{s}}
 \altaffiliation{Present address: SLAC National Accelerator Lab., 2575 Sand Hill Rd, Menlo Park, CA 94025, USA}
 \affiliation{Max Planck Institute for Chemical Physics of Solids, N{\"o}thnitzer Stra{\ss}e 40, 01187 Dresden, Germany}
 \affiliation{PETRA III, Deutsches Elektronen-Synchrotron (DESY), Notkestra{\ss}e 85, 22607 Hamburg, Germany}
\author{Andrei~Gloskovskii}
 \affiliation{PETRA III, Deutsches Elektronen-Synchrotron (DESY), Notkestra{\ss}e 85, 22607 Hamburg, Germany}
\author{Christoph~Schlueter}
 \affiliation{PETRA III, Deutsches Elektronen-Synchrotron (DESY), Notkestra{\ss}e 85, 22607 Hamburg, Germany}
\author{Sheng Ran}
  \altaffiliation{Present address: Department of Physics, University of Maryland, College Park, MD 20742, USA;  NIST Center for Neutron Research, National Institute of Standards and Technology, Gaithersburg, MD 20899, USA}
  \affiliation{Department of Physics, University of California, San Diego, La Jolla, California, USA}
\author{M. Brian Maple}
  \affiliation{Department of Physics, University of California, San Diego, La Jolla, California, USA}
\author{Peter~Thalmeier}
	\affiliation{Max Planck Institute for Chemical Physics of Solids, N{\"o}thnitzer Stra{\ss}e 40, 01187 Dresden, Germany}
\author{Liu~Hao~Tjeng}
	\affiliation{Max Planck Institute for Chemical Physics of Solids, N{\"o}thnitzer Stra{\ss}e 40, 01187 Dresden, Germany}
\author{Andrea~Severing}
  \affiliation{Institute of Physics II, University of Cologne, Z{\"u}lpicher Stra{\ss}e 77, 50937 Cologne, Germany}
  \affiliation{Max Planck Institute for Chemical Physics of Solids, N{\"o}thnitzer Stra{\ss}e 40, 01187 Dresden, Germany}	
\date{\today}

\begin{abstract}
The application of pressure as well as the successive substitution of Ru with Fe in the hidden order (HO) compound URu$_2$Si$_2$ leads to the formation of the large moment antiferromagnetic phase (LMAFM). Here we have investigated the substitution series URu$_{2-x}$Fe$_x$Si$_2$ with $x$\,=\,0.2 and 0.3 with non-resonant inelastic x-ray scattering (NIXS) and 4$f$ core-level photoelectron spectroscopy with hard x-rays (HAXPES). NIXS shows that the substitution of Fe has no impact on the symmetry of the ground-state wave function. In HAXPES we find no shift of spectral weight that would be indicative for a change of the 5$f$-electron count. Consequently, changes in the exchange interaction $\cal{J}$ due to substitution must be minor so that the conjecture of chemical pressure seems unlikely. An alternative scenario is discussed, namely the formation of long range magnetic order due the substitution induced local enhancement of the magnetization in the vicinity of the $f$-electron ions while the overall electronic structure remains unchanged. 

\end{abstract}

\maketitle
\section{Introduction}
The transition into an electronically ordered state at 17.5\,K in the heavy fermion compound URu$_2$Si$_2$ has attracted an enormous amount of interest since its discovery about 35 years ago\,\cite{Palstra1985,Schlabitz1986,Maple1986,Oppeneer2010,Mydosh2011,Mydosh2014,Mydosh2020}, and yet, the order parameter of this phase is still a matter of debate\,\cite{Ikeda2012,Kung2015}. The small antiferromagnetic ordered moment of 0.03\,$\mu_B$/U along the tetragonal $c$ axis\,\cite{Broholm1987,Niklowitz2010} is too small to account for the loss of  entropy of about 0.2$R$ln2 and changes in transport properties so that the presence of long-range magnetic order, charge density or spin density wave order can be excluded and the name \textit{hidden order} (HO) phase was born. At about 1.5\,K, URu$_2$Si$_2$ undergoes a second transition into an unconventional superconducting state. 

In heavy fermion compounds the exchange interaction $\cal{J}$\,$\approx$\,$V^2$/$\epsilon_f$, with $V$ the hybridization strength of $f$ and conduction electrons and -$\epsilon_f$ the $f$-level position relative to the Fermi level plays a crucial role in the ground-state formation\,\cite{Floquet2005,Thalmeier2005,Coleman2007,Hilbert2007,Khomskii2010,Stockert2012,White2015}. How to formulate this process is, however, a subject of intense discussions. Band effects are clearly important so band structure approaches\,\cite{Oppeneer2010,Mydosh2011,Mydosh2014} have their merits. Yet a localized 5$f$ electron picture may also have its value. Recently, the existence of local atomic multiplet states has been observed for URu$_2$Si$_2$ and the symmetry of the local ground-state wave function can be well described by a singlet state or quasi-doublet consisting of two singlets that belong to the U$^{4+}$ 5$f^2$ configuration\,\cite{Sundermann2016}. The interplay with the bands are then represented by the non-integer filling of the 5$f$ shell\,\cite{manuscript2020}, i.e. more than one configuration contributes to the ground state.

\begin{figure}[]
    \includegraphics[width=0.90\columnwidth]{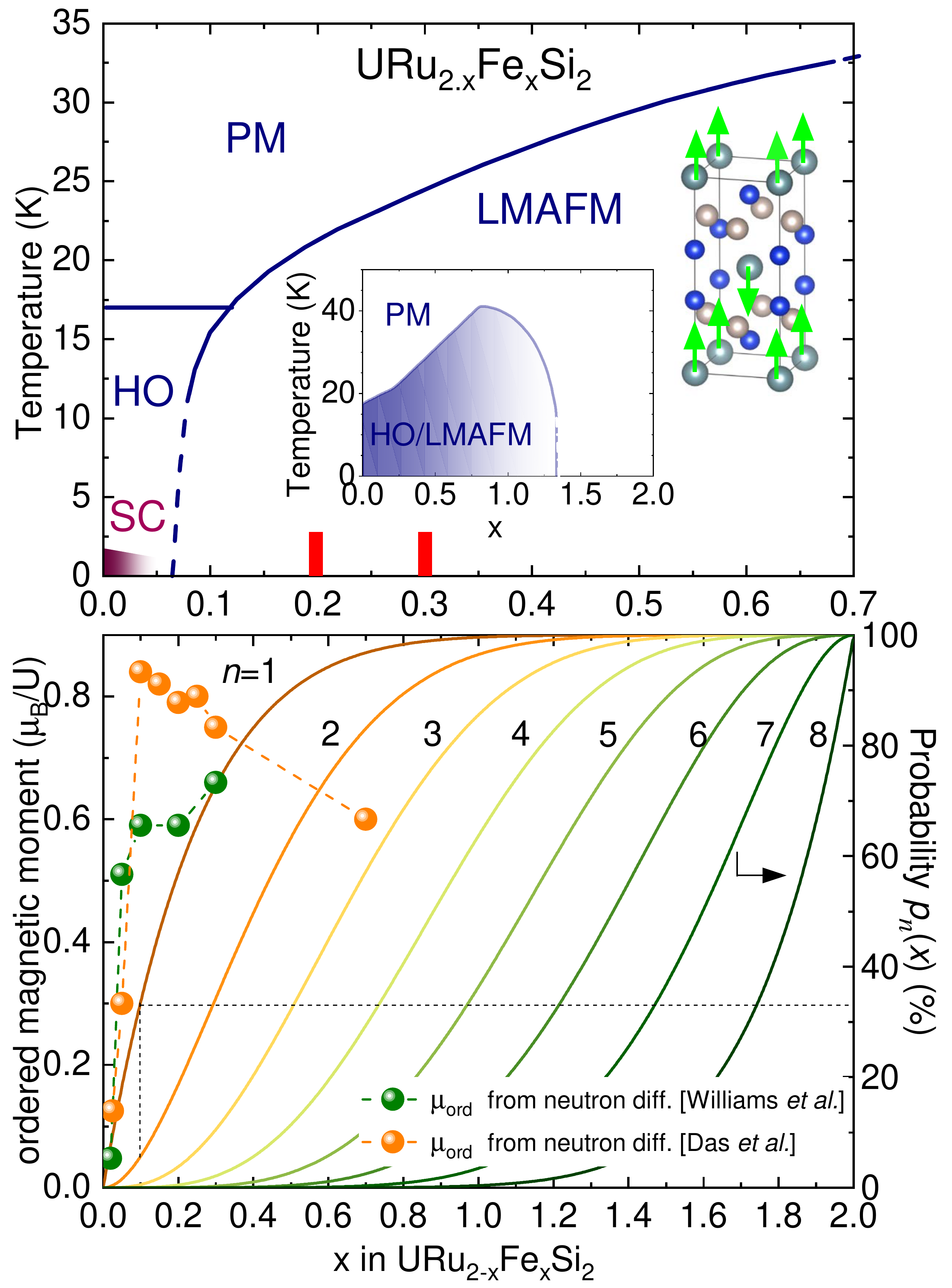}
    \caption{(color online) (a) Temperature versus $x$, $x$\,=\,Fe concentration, phase diagram adapted from Ref.\,\onlinecite{Ran2016} and  \onlinecite{Kanchanavatee2011} showing the phase boundaries from the paramagnetic phase (PP) to the large moment antiferromagnetic (LMAFM) and superconducting (SC) phase. The red ticks mark the concentrations used in HAXPES. The cartoon of the crystal lattice shows the tetragonal unit cell of URu$_{2-x}$Fe$_x$Si$_2$ (U silver, Fe/Ru gold, Si blue), the green arrows symbolize the antiferromagnetic structure in the LMAFM phase; (b) ordered magnetic moments (left scale), orange dots adapted from Ref.\,\onlinecite{Das2015}, green dots adapted from Ref.\,\onlinecite{Williams2017}, plotted on top of the  probability functions $p_n(x)$ for finding U surrounded by at least $n$\,=\,1,\,2,\,3,\,4, up to 8 Fe ions as function of Fe concentration $x$ (colored lines, right scale)\,\cite{Jaccarino1965}. }
    \label{prob}
\end{figure}

The application of pressure suppresses the HO phase and a large moment antiferromagnetic (LMAFM) phase develops. The ordered magnetic moments are aligned along the tetragonal $c$ axis (see insest of Fig.\,\ref{prob}\,(a)) and its value rises discontinuously from 0.03 to about 0.4\,$\mu_B$ at a critical pressure of about 5\,kbar \,\cite{Amitsuka2007,Butch2010,Niklowitz2010,Das2015,Williams2017} while the transition temperature ($\approx$20\,K at $p$\,=\,15\,kbar) rises only slightly. The Fermi surfaces of the HO and LMAFM phase are very similar according to Shubnikov–de Haas measurements\,\cite{Hassinger2010} and the doubling of the unit cells seemingly already takes place in the HO state.
  
The substitution of Fe on the Ru site also leads to the formation of antiferromagnetic order\,\cite{Kanchanavatee2011,Das2015,Wilson2016,Butch2016,Ran2016,Wolowiec2016,Kung2016,Ran2017,Williams2017,Kissin2019}. It has been interpreted as the possible effect of chemical pressure by the smaller ionic radius of Fe and indeed, the phase diagram of URu$_{2-x}$Fe$_x$Si$_2$, temperature $T$ versus Fe amount $x$ (see Fig.\,\ref{prob}\,(a)), bears similarities to the $T$-$p$ phase diagram of URu$_2$Si$_2$\,\cite{Das2015,Kanchanavatee2011}. Neutron diffraction shows magnetic moments of about 0.05 to 0.1\,$\mu_B$ for $x$\,$\approx$\,0.025 and the ordered magnetic moment $\mu_{\text{ord}}$ rises quickly with $x$\,\cite{Das2015} and already reaches its maximum value of about 0.8\,$\mu_B$ for $x$\,=\,0.1, according to Ref.\,\onlinecite{Das2015}. In this work the Fe concentration was checked with EDX (see orange dots in Fig.\,\ref{prob}\,(b)). Slightly smaller moments have been observed by the authors of Ref.\,\onlinecite{Williams2017} (see green dots in Fig.\,\ref{prob}\,(b)). A possible reason for this discrepancy could be differences in sample stoichiometry. With further increase of $x$, the magnetic moment decreases while the ordering temperature continues to increase up to $x$\,=0.8 and then quickly drops and reaches zero at about $x$\,=\,1.2\,\cite{Kanchanavatee2011} (see inset fo Fig\,\ref{prob}\,(a)).  For even larger $x$, the system adopts a Pauli paramagnetic state as UFe$_2$Si$_2$. The lattice constant of the long tetragonal $c$ axis remains almost unchanged with increasing Fe concentration, whereas the short $a$ axis decreases linearly from $x$\,=\,0 to $x$\,=\,2\,\cite{Kanchanavatee2011}. The magnetic volume fraction has been determined with $\mu$SR and amounts to 0.6 for $x$\,=\,0.02 and 1 for $x$\,$\geq$\,0.1\,\cite{Wilson2016}. 

Here we want to question the conjecture that chemical pressure drives URu$_{2-x}$Fe$_x$Si$_2$ into an antiferromagnetic state. Chemical pressure upon Fe substitution does not explain that small amounts of substitution with Os have the same effect\,\cite{Dalichaouch1990,Kanchanavatee2014,Wilson2016} and also does not explain the suppression of the LMAFM phase for higher Fe concentrations. The appearance of magnetic order for the smaller Fe concentration is also puzzling in view of UFe$_2$Si$_2$ exhibiting enhanced Pauli paramagnetism (PP) down to the lowest measured temperature\,\cite{Szytuka1988,Szytuka2007,Endstra1993b}. Hence, the appearance of antiferromagnetic (AFM) order upon Fe susbstitution should be treated as a great surprise. 

We therefore investigated the electronic structure of URu$_{2-x}$Fe$_x$Si$_2$ with $x$\,=\,0.2 and 0.3. For both concentrations URu$_{2-x}$Fe$_x$Si$_2$ is well placed in the antiferromagnetic region (LMAFM) of the $T$-$x$ phase diagram (see red ticks in Fig.\,\ref{prob}\,(a)). We probed the impact of the Fe concentration $x$ on the ground-state symmetry with non-resonant inelastic x-ray scattering (NIXS) and the $relative$ filling of the U 5$f$ shell with hard x-ray photoelectron spectroscopy (HAXPES). Any changes in the 5$f$ electron count with $x$ would then point towards a change in the exchange interaction $\cal{J}$ but only if the symmetry remains unaffected by substitution. 

The ground state symmetry in the U$M_2$Si$_2$ family with I4/mmm structure is given by one of the seven crystal-field states of the Hund's rule ground state of U$^{4+}$\,5$f^2$ with $J$\,=\,4. It turned out that the different ground state properties of the U$M_2$Si$_2$ ($M$\,=\,Ni, Pd: AFM, $M$\,=\,Ru: HO; $M$\,=\,Fe: PP) arise out of the same crystal-field symmetry, namely a singlet or a quasi-doublet state consisting of the $\Gamma_1^{(1)}$ with strong $J_z$\,=\,$+4$ and $-4$ contributions and the $\Gamma_2$\,\cite{Sundermann2016,manuscript2020}. Note, only the quasi-doublet gives rise to a sizable ordered magnetic moment. Having the same symmetry, the relative 5$f$ electron count can be extracted in a straightforward manner from the HAXPES spectra. The 4$f$ core-level data show a clear shift of spectral weights from U$^{3+}$\,5$f^3$ to U$^{4+}$\,5$f^2$ when comparing the data of compounds with a PP, HO and AFM collectively ordered ground states\,\cite{Sundermann2016,manuscript2020}. Following this line of thought, we expect a decrease of U$^{3+}$\,5$f^3$ spectral weight for Fe substitutions in the LMAFM phase if the Fe substitution changes the exchange interaction of the U 5$f$ and conduction electrons.  

We present 4$f$ core-level HAXPES data of the substitution series URu$_{2-x}$Fe$_x$Si$_2$ with $x$\,=\,0, 0.2 and 0.3 and compare the results with 4$f$ core-level data of URu$_2$Si$_2$ (HO). We are searching for relative changes in the U 5$f$-shell occupations with the Fe concentration $x$. First of all, however, we have to verify that the ground state symmetry remains unchanged upon Fe substitution so that we show non-resonant inelastic x-ray scattering (NIXS) data of the sample with the highest Fe concentration (URu$_{1.7}$Fe$_{0.3}$Si$_2$).

\section{Experiment}
Single crystals of Fe-substituted URu$_2$Si$_2$ were grown by the Czochralski method in a tetra-arc furnace from high purity starting elements (depleted uranium\,-\,3N, Ru, Fe\,- 3N, Si\,- 6N). Real Fe concentration was examined by elemental analysis using energy dispersive x-ray spectroscopy, which is uniform throughout the sample and in a good agreement with the norminal concentration.  

The NIXS experiment was performed at the High-Resolution Dynamics Beamline P01 of the PETRA-III synchrotron in Hamburg, Germany. The end station has a vertical geometry with twelve Si(660) 1\,m radius spherically bent crystal analyzers that are arranged in a 3$\times$4 matrix and positioned at scattering angles of 2\,$\theta$\,$\approx$\,150$^\circ$, 155$^\circ$, and 160$^\circ$. The final energy was fixed at 9690\,eV, the incident energy was selected with a Si(311) double monochromator, and the overall energy resolution was $\approx$\,0.7\,eV. The scattered beam was detected by a position sensitive custom-made Lambda detector based on a Medipix3 chip. More details about the experimental set-up can be found in Ref.\,\cite{Sundermann2017}. The averaged momentum transfer was $|\vec{q}|$\,=\,(9.6\,$\pm$\,0.1)\,\AA$^{-1}$ at the U $O_{4,5}$ edge. The sample was mounted in a Dynaflow He flow cryostat with Al-Kapton windows and the temperature was set to 15\,K.

The HAXPES experiments were carried out at beamline P09 of the PETRA-III synchrotron in Hamburg, Germany\,\cite{Strempfer2013}. The incident photon energy was set at 5945\,eV. The valence band spectrum of a gold sample was measured in order to determine the Fermi level $E_F$ and the overall instrumental resolution of 300\,meV. The excited photoelectrons were collected using a SPECS225HV electron energy analyzer in the horizontal plane at 90$^{\circ}$. The sample emission angle was 45$^{\circ}$. Clean sample surfaces were obtained by cleaving the samples \textsl{in situ} in the cleaving chamber prior to inserting them into the main chamber where the pressure was $\sim$10$^{-10}$\,mbar. The measurements were performed at a temperature of 20\,K. 


\section{Results}
\begin{figure}[t]
    \includegraphics[width=0.95\columnwidth]{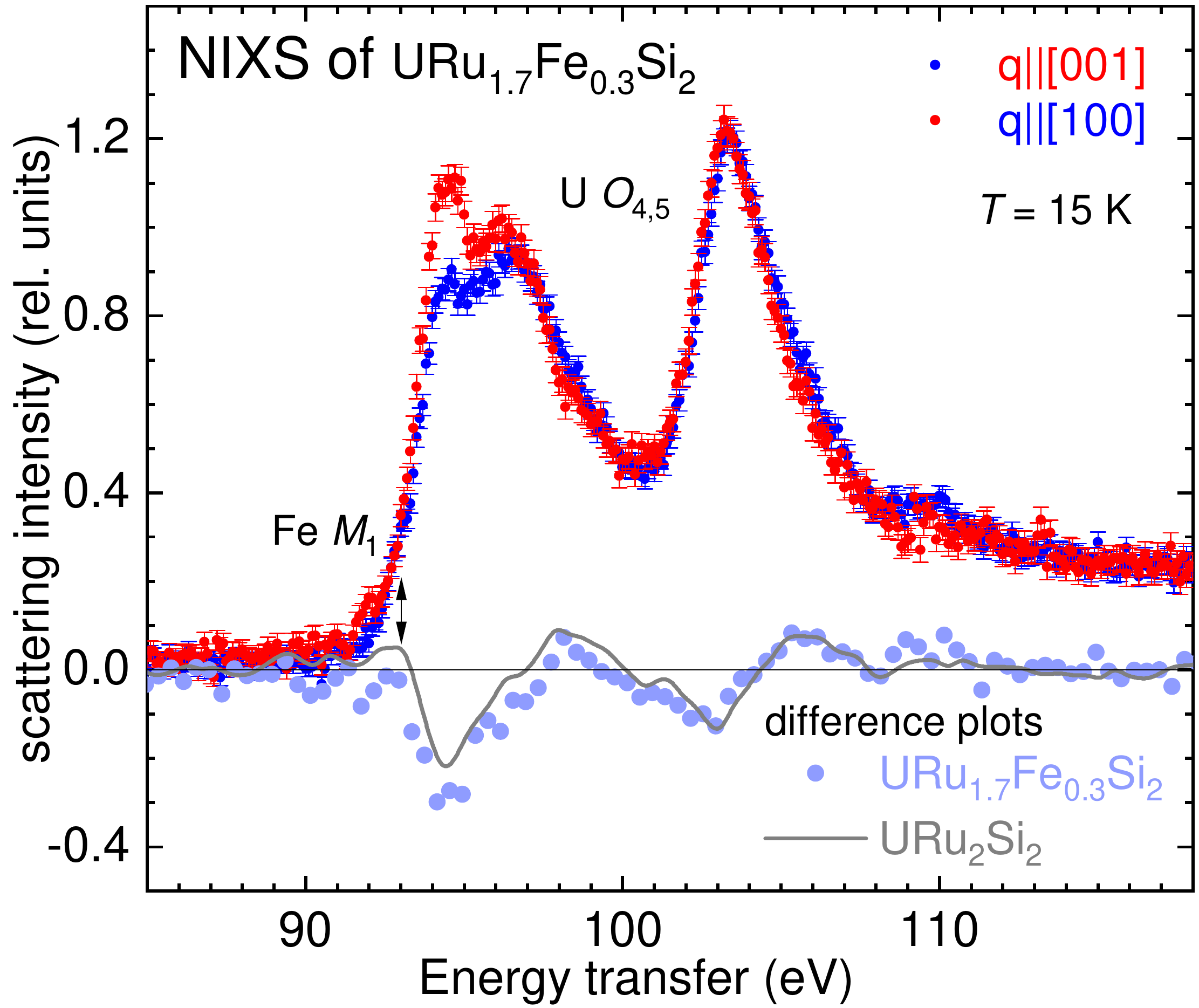}
    \caption{(color online) Normalized and background corrected experimental NIXS data of URu$_{1.7}$Fe$_{0.3}$Si$_2$ at the U $O_{4,5}$ edges (5$d$\,$\rightarrow$\,5$f$) at $T$\,=\,15\,K for $\vec{q}$$\|$[100] (blue dots) and  $\vec{q}$$\|$[001] (red dots), plus the difference plots I$_{\vec{q}\|[100]}$-I$_{\vec{q}\|[001]}$ (violet dots). For comparison the difference data of URu$_2$Si$_2$ are shown (gray line), adapted from Ref.\,\cite{Sundermann2016}. When no error bars are given the size of the data points represent the statistical error.}
    \label{NIXS}
\end{figure}

Figure\,\ref{NIXS} shows the U $O_{4,5}$ edge (5$d$\,$\rightarrow$\,5$f$) NIXS data of URu$_{1.7}$Fe$_{0.3}$Si$_2$ at 15\,K for two directions of the momentum transfer $\vec{q}$, for $\vec{q}$$\|$[100] (blue) and $\vec{q}$$\|$[001] (red). Two aspects are striking, namely the strong directional dependence and the existence of a multiplet structure\,\cite{Sundermann2016,manuscript2020}. In NIXS, the scattering signal depends on the direction of momentum transfer in the same way as in x-ray absorption spectroscopy (XAS), the signal depends on the direction of the electric field vector of the linear polarized light. In analogy to XAS\,\cite{Hansmann2008,Willers2015}, NIXS\,\cite{Willers2012,Sundermann2016,Sundermann2017,Sundermann2018,Sundermann2018a} therefore gives insight into the orbital occupation. XAS and NIXS both follow selection rules; in the case of XAS, dipole selection rules govern the transitions probabilities\,\cite{deGroot1994}, in the case of NIXS at large $|\vec{q}|$, so called multipole selection rules are in place\,\cite{Sundermann2017,Gordon2009,Bradley2010,Caciuffo2010,Gupta2011,Willers2012,Sundermann2018}. The appearance of the  multiplet structure in contrast to the broad Fano-like lineshape of the U $O_{4,5}$ edge in XAS is due to the more excitonic character of the NIXS spectra at large $|\vec{q}|$ (compare e.g. \cite{Wray2015}).

At the bottom of Fig.\,\ref{NIXS}, the difference plot I$_{\vec{q}\|[100]}$-I$_{\vec{q}\|[001]}$ is shown (violet dots) and compared to the difference spectrum of URu$_2$Si$_2$ (gray line). The URu$_2$Si$_2$ data are adapted from Ref.\,\cite{Sundermann2016}. The directional dependent signal of the pure Ru and of the 15\% Fe substituted sample are almost the same. Only in the energy region around 95\,eV energy transfer is the directional signal of the Fe substituted sample slightly larger, but this is due to the contribution of the dipole forbidden Fe\,$M_1$ core level. It contributes to the scattering intensity and also to the directional dependence\,\cite{manuscript2020}. Above 100\,eV energy transfer, where the U $O_{4,5}$ edge signal is free of Fe scattering, the dichroisms of URu$_{1.7}$Fe$_{0.3}$Si$_2$ and URu$_2$Si$_2$ are identical so that we also conclude the ground-state symmetries are described by the same states, namely the $\Gamma_1^{(1)}$ or the $\Gamma_2$ singlet or a quasi-doublet consisting of these two states. The other five crystal-field states of the U$^{4+}$\,5$f^2$ configuration with $J$\,=\,4 in tetragonal point symmetry exhibit either a much smaller dichroism or a dichroism with the opposite sign\,\cite{Sundermann2016,manuscript2020}.


\begin{figure}[]
    \includegraphics[width=0.95\columnwidth]{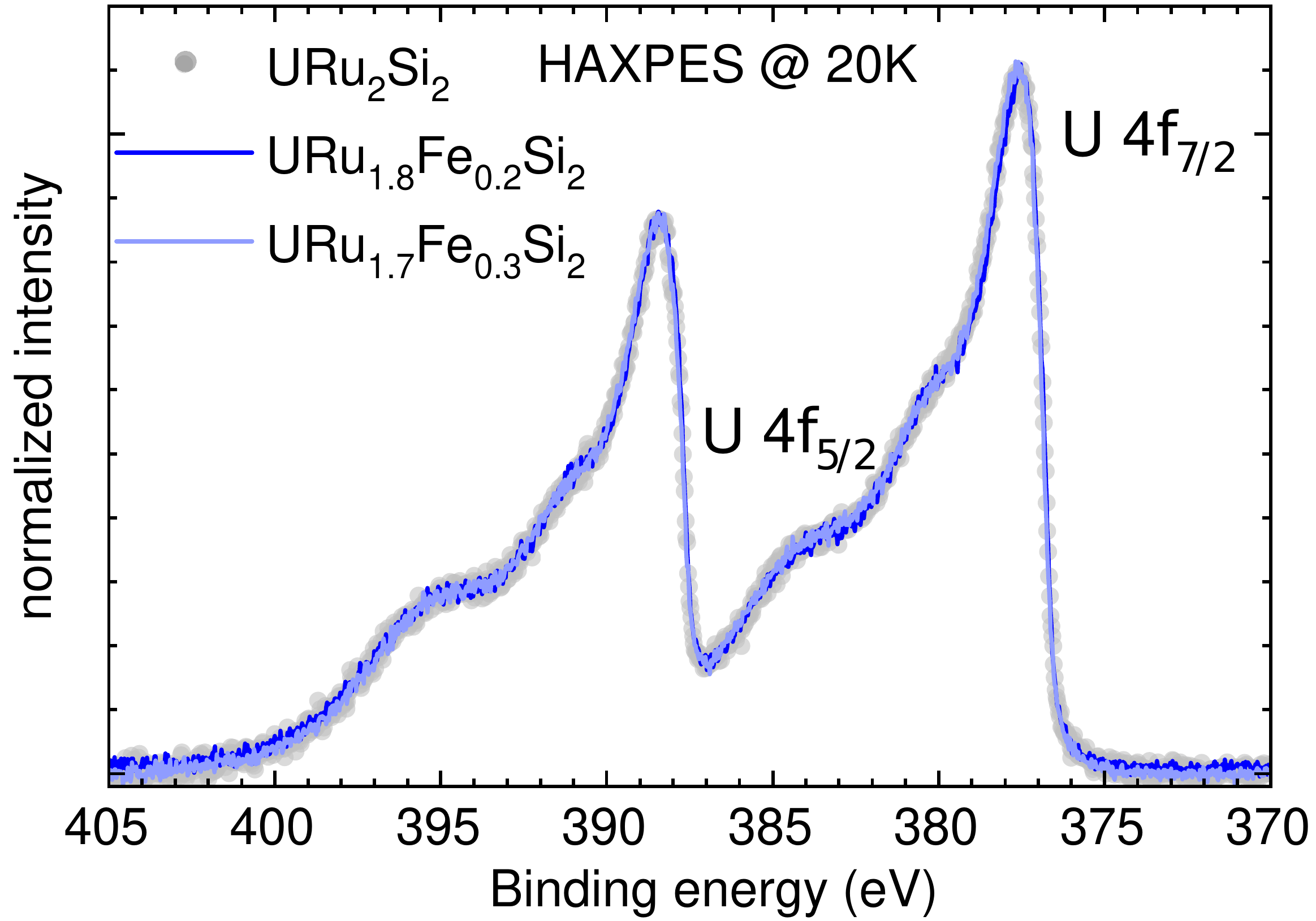}
    \caption{(color online) U 4$f$ core level data of URu$_{2-x}$Fe$_x$Si$_2$ with x\,=\,0.2 and 0.3 (blue lines) and of URu$_2$Si$_2$ (gray dots) for comparison. The URu$_2$Si$_2$ data are adpated from Ref.\,\cite{manuscript2020}. All data are background corrected and normalized to the integrated intensity.}
    \label{4f}
\end{figure}

Figure\,\ref{4f} shows the U\,4$f$ core-level spectra of the two URu$_{2-x}$Fe$_x$Si$_2$ samples (blue-purple lines) after the subtraction of an integrated (Shirley-type) background and normalization to the the integrated intensity, superimposed on the data of URu$_2$Si$_2$ (gray dots) which are adapted from Ref.\,\cite{manuscript2020}. We find that there is no difference between the core-level spectra of the substitution series despite the different ground state properties. We therefore conclude that the 5$f$ shell occupation does not change or rather, it changes so little due to Fe substitution up to x\,=\,0.3 that it is below the detection limit.

\section{Discussion}
The above NIXS results shows that the Fe substituted samples in the LMAFM phase have the same crystal-field ground-state symmetry as the HO compound URu$_2$Si$_2$. Hence, the 4$f$ core-level HAXPES data of the substitution series are directly comparable. The U4$f$ core-level HAXPES spectra do not exhibit any shift of spectral weight upon substitution with Fe. Neither a decrease nor increase has been detected although strong shifts of spectral weights were observed when comparing antiferromagnetic UPd$_2$Si$_2$ and UNi$_2$Si$_2$, the hidden order compound URu$_2$Si$_2$, and the Pauli paramagnet UFe$_2$Si$_2$\,\cite{manuscript2020}.  This implies that the 5$f$ shell occupation and with it the exchange interaction $\cal{J}$ does not change for Fe concentrations of up to $x$\,=\,0.3. We may further conclude that also the hybridization $V$ remains unaffected by these small concentrations of Fe unless changes in $V^2$ and $\epsilon_f$ cancel each other out. This finding contradicts the chemical pressure argument leading to antiferromagnetic order in URu$_{2-x}$Fe$_x$Si$_2$. In addition, the aforementioned puzzle that small Os substitutions also lead to antiferromagnetic order and that UFe$_2$Si$_2$ is a Pauli paramagnetic suggest that the appearance of magnetic order in the lower doping regime of URu$_{2-x}$Fe$_x$Si$_2$ must have a different cause than chemical pressure.

The NIXS results of the same isostructural U$M_2$Si$_2$ family ($M$\,=\,Fe, Ni, Ru, Pd) are compatible with two singlet states close in energy forming a quasi-doublet ground state so that $induced$ order has been suggested to be the most likely mechanism for the formation of the antiferromagnetic ground states in UPd$_2$Si$_2$ and UNi$_2$Si$_2$\,\cite{manuscript2020}. Furthermore, for the example of UPd$_2$Al$_3$, it had been shown that in case of induced order, the N\`eel temperature as well as the size of the ordered magnetic moments depend on two parameters\,\cite{Thalmeier2002}; the size of the energy splitting within the quasi-doublet and the strength of the exchange interaction $\cal{J}$. What does this imply for URu$_{2-x}$Fe$_x$Si$_2$? In HAXPES, a change of spectral weights is beyond detection which suggests that $\cal{J}$ remains next to unchanged for small substitutions of Fe up to $x$\,=\,0.3. However, the local crystal-field could change, e.g. by local distortions due to the smaller ionic radius of Fe. It would then be imaginable that locally a moment is induced if the splitting of the two singlet states forming the quasi-doublet decreases with $x$.

Already in 1962, Jaccarino and Walker discussed the appearance of magnetization for small impurities in metallic hosts in terms of the neighboring configurations of ions in alloy systems\,\cite{Jaccarino1965}. Figure\,\ref{prob}\,(b) shows the probability $p_n$($x$) for the U atoms to be surrounded by at least $n$\,=\,1,\,2, up to 8 Fe neighbors as function of the amount of Fe substitution $x$. We have superimposed the ordered magnetic moments as measured with neutron diffraction and, although there are some differences in the two data sets, it is apparent that less than one Fe neighbor per unit cell is sufficient to cause magnetic order and to reach the maximum moment. Furthermore, for a substitution level of $x$\,=\,0.1, i.e., already well in the LMAFM regime, only about 30\% of the U ions have one Fe ion as a direct neighbor (see dashed lines in Fig.\,\ref{prob}\,(b)). We now follow 
 the idea of Sakai \textsl{et al.} for the formation of magnetic order in the heavy fermion superconductor CeCoIn$_5$ upon Cd doping\,\cite{Sakai2015} in order to explain how  magnetic order can appear for very small amounts of substitution. Upon substitution, the small local change in the electronic states induces local spins on neighboring U (Ce) sites while the majority of the electronic states remain unchanged. The $nucleation$ of short-range ordering near the Fe (Cd) dopants leads to long-range AFM ordering. 
 
With further increase of the Fe concentration, the impact of stronger U\,5$f$-Fe\,3$d$ exchange interaction will gain weight so that eventually magnetic order breaks down and a Pauli paramagnetic state forms. It is intriguing that magnetic order breaks down just above $x$\,$\approx$\,$\geq$\,1 (see inset of Fig.\,\ref{prob}\,(a)\,\cite{Kanchanavatee2011}); i.e., when the majority of U atom is surrounded by more than 4 Fe neighbors so that the environment of U is Fe and no longer Ru dominated. Then, seemingly, the stronger exchange interaction of Fe and U with respect to U and Ru determines the ground state.

\section{Conclusion}
The substitution series URu$_{2-x}$Fe$_x$Si$_2$ with $x$\,=\,0.2 and 0.3 has been investigated with non-resonant inelastic x-ray scattering (NIXS) and hard x-ray 4$f$ core-level photoelectron spectroscopy (HAXPES). The crystal-field symmetry of the ground state remains unchanged upon substitution with Fe so that the U 4$f$ core-level HAXPES data are directly comparable. HAXPES reveals no shift of spectral weights upon Fe substitution thus making it unlikely that changes in the exchange interaction are responsible for the formation of large moments. The combination of the present spectroscopic findings and the fact that already less than one Fe ion surrounding U is sufficient for the formation of sizable moments makes the scenario of Fe substitution being analogous to chemical pressure unlikely.


\section{Acknowledgment} This research was carried out at PETRA\,III/DESY, a member of the Helmholtz Association HGF. Research at UC San Diego was supported by the US Department of Energy, Office of Basic Energy Sciences, Division of Materials Science and Engineering, under Grant No. DEFG02-04-ER45105 (single crystal growth) and US National Science Foundation under Grant No. DMR-1810310 (materials characterization). A.A., A.S., and M.S. gratefully acknowledge the financial support of the Deutsche Forschungsgemeinschaft under project SE\,1441-5-1.

\end{document}